\documentclass[aps,pra,reprint,showpacs,showkeys,groupedaddress]{revtex4-1}
\usepackage{amsmath}
\begin{document}

%\preprint{}

\title{Quantum entropy-typical subspace and universal data compression}


\author{Jingliang Gao}
\email[]{gaojl0518@gmail.com}
\author{Yanbo Yang}
%\thanks{}
\affiliation{State Key Laboratory of Integrated Services Networks, Xidian University, Xi'an 710071, China}


\date{April 2, 2014}

\begin{abstract}
The quantum entropy-typical subspace theory is specified. It is shown that any $\rho^{\otimes n}$ with von Neumann entropy$\leq h$ can be preserved approximately by the entropy-typical subspace with entropy$=h$. This result implies an universal compression scheme for the case that the von Neumann entropy of the source does not exceed $h$.
\end{abstract}


\maketitle

%\subsection{}
%\subsubsection{}
Quantum data compression is one of the most fundamental tasks in quantum information theory\cite{Nielsen2000,Wilde2011}. Schumacher first provided a tight bound (equal to the von Neumann entropy of the source) to which quantum information may be compressed\cite{Schmacher1995}. From then on, many compression schemes have been proposed\cite{Koashi2001,Bostroem2002,Jozsa2003,Hayashi2010,Kaltchenko2003,
Bennett2006,Jozsa1998,Hayashi2002a,Hayashi2002b,Braunstein2000}. In this paper we will consider the universal quantum data compression in the case that we only know the entropy of the source does not exceed some given value h. In classical information, an explicit example of such compression is a scheme based on the theory of types developed by Csiszar and K\"{o}rner\cite{CK2011}. They showed that the data can be compressed to $h$ bits/siganl by encoding all the sequences $x^{n}$ for which $H(\mathbf {p_{x}})\leq h +\varepsilon$(called $CK$ sequence), where $\mathbf{p_{x}}$ is the type of $x^{n}$ and $H(\cdot)$ is the Shannon entropy function. For quantum information, an analogous theory was established\cite{Jozsa1998} by Jozsa et al. They extended the classical $CK$ sequence to a quantum subspace $\Xi(B)$ for a given basis $B$, and then to $ \Upsilon$ which is the span of $ \Xi(B)$ as $B$ ranges over all bases. They proved that the dimension of $\Upsilon$ is up to some polynomial multiple of $\dim{\Xi(B)}$, so the compression rate h is achievable asymptotically. We note that, their proof is based on the $CK$ sequence set, so a natural question arises: Is the $CK$ set essential to the proof? Or can it be replaced by a smaller set? In this paper, we give the answer. It will be shown that, if we replace the $CK$ set with the entropy-typical set $\{ x^{n}:\left| H(\mathbf{p_{x}})-h \right|\leq \varepsilon \} $, the proof still holds. This result is based on the quantum entropy-typical subspace theory which reveals that
any $\rho^{\otimes n}$ with entropy$\leq h$ can be preserved by the entropy-typical subspace with entropy$=h$.

Before presenting our main results, we begin with describing some basic concepts which will be used later.

Let $\chi=\{ 1, 2,..., d\}$ be a alphabet with $d$ symbols. We use $\mathbf{p}=\left( p(1), p(2),...,p(d) \right)$ to denote a probability distribution on $\chi$, where $p(a)$ is the probability of the symbol $a$. Let $X_{1}, X_{2},..., X_{n}$ be a sequence of n symbols from $\chi$. We will use the notation $x^{n}$ to denote a sequence $x_{1}, x_{2},..., x_{n}$.

The type $\mathbf{p_{x}}$ of $x^{n}$ is the relative proportion of occurrences of each symbol in $\chi $, i.e. $\mathbf{p_{x}}(a)=N(a|x^{n})/n$ for all $ a\in \chi$, where $N(a|x^{n})$ is the number of times the symbol $a$ occurs in the sequence $x^{n}$.

The strongly typical set of a source with distribution $\mathbf {p}$ is defined as:
\begin{eqnarray}
A_{\varepsilon}(\mathbf{p})= \bigg \{ x^{n}: \left| \mathbf{p_{x}}(a)-p(a) \right| \leq  \frac{\varepsilon}{\left|\chi\right|}, \forall \  a\in \chi \bigg \}
\label{eqn:1}
\end{eqnarray}
$A_{\varepsilon}(\mathbf{p})$ is a high probability set\cite{Cover2012}, i.e. for any fixed $\varepsilon > 0$ and $\delta > 0$, when n is large enough,
\begin{eqnarray}
\sum_{x^{n} \in A_{\varepsilon}(\mathbf{p})}p(x^{n}) \geq 1-\delta
\label{eqn:2}
\end{eqnarray}
where $p(x^{n})=p(x_{1})p(x_{2})...p(x_{n})$.

% If in two-column mode, this environment will change to single-column
% format so that long equations can be displayed. Use
%\begin{widetext}
%\end{widetext}
Now we specify the definition of classical entropy-typical set.

% figures should be put into the text as floats.
% Use the graphics or graphicx packages (distributed with LaTeX2e)
% and the \includegraphics macro defined in those packages.
% See the LaTeX Graphics Companion by Michel Goosens, Sebastian Rahtz,
\noindent\textbf{Definition1:} \emph{Given $\varepsilon > 0$ and $h > 0$, the classical entropy-typical set $T_{\varepsilon}(h)$ is defined as:}
\begin{eqnarray}
T_{\varepsilon}(h)=\{ x^{n}:\left| H(\mathbf{p_{x}})-h \right|\leq \varepsilon \}
\label{eqn:3}
\end{eqnarray}
\emph{where $H(\cdot)$ is the Shannon entropy function.}\\

\noindent\emph{Property1.1.} According to the type method theory\cite{CK2011,Cover2012}, we can easily know, for any $\varepsilon >0$,
\begin{eqnarray}
\left | T_{\varepsilon}(h) \right | \leq  (n+1)^{d} 2^{n(h+\varepsilon)}
\label{eqn:4}
\end{eqnarray}
\noindent\emph{Property1.2.} For any source with $H(\mathbf{p})=h $, $T_{\varepsilon}(h)$ is a high-probability set, i.e. for any $\varepsilon > 0$ and $\delta > 0$, for sufficiently large n,
\begin{eqnarray}
\sum_{x^{n} \in T_{\varepsilon}(h)}p(x^{n}) \geq 1-\delta
\label{eqn:5}
\end{eqnarray}
\emph{Proof}: It is easy to show that $T_{\varepsilon}(h)$ contains a subset $A_{\varepsilon^{\prime}}(\mathbf{p})$. Since $H(\cdot)$ is a continuous function, for any $\varepsilon >0$ there exist $\varepsilon^{\prime} >0$ such that $\left| H(\mathbf{p_{x}})-H(\mathbf{p})\right| \leq \varepsilon$ for all $\left| \mathbf{p_{x}}-\mathbf{p} \right| \leq  \varepsilon^{\prime}$. Combined with the definition of the strongly typical set, we see that, $\left| H(\mathbf{p_{x}})-H(\mathbf{p})\right|=\left| H(\mathbf{p_{x}})-h\right| \leq \varepsilon$ holds for all $x^{n}\in A_{\varepsilon^{\prime}}(\mathbf{p}) $, which means $A_{\varepsilon^{\prime}}(\mathbf{p})\subseteq T_{\varepsilon}(h)$.
Thus for sufficiently large n,
\begin{eqnarray}
\sum_{x^{n} \in T_{\varepsilon}(h)}p(x^{n}) \geq \sum_{x^{n} \in A_{\varepsilon^{\prime}}(\mathbf{p})}p(x^{n}) \geq 1-\delta
\label{eqn:6}
\end{eqnarray}
\\

Let $\mathcal{H}$ be a d-dimensional Hilbert space and  $B=\left \{ \left |  e_{1} \right \rangle,\left |  e_{2} \right \rangle,...,\left |  e_{d} \right \rangle \right \}$ be a basis of $\mathcal{H}$. We can extend $T_{\varepsilon}(h)$ to quantum case.\\

\noindent\textbf{Definition2.} \emph{The entropy-typical subspace for a given basis B can be defined as:}
\begin{eqnarray}
\Xi(h,B)=span\{ \left | e_{x_{1}}e_{x_{2}}...e_{x_{n}} \right \rangle:x^{n} \in T_{\varepsilon}(h) \}
\label{eqn:7}
\end{eqnarray}
Denote the projector onto $\Xi(h,B)$ by $\Pi(h,B)$,
\begin{eqnarray}
\Pi(h,B)=\sum_{x^{n} \in T_{\varepsilon}(h)}\ | e_{x_{1}}e_{x_{2}}...e_{x_{n}} \rangle \langle e_{x_{1}}e_{x_{2}}...e_{x_{n}} \ |
\label{eqn:8}
\end{eqnarray}
From the properties of $T_{\varepsilon}(h)$, we can get the properties of $\Xi(h,B)$.\\
\emph{Property2.1.} For any $\varepsilon > 0$,
\begin{eqnarray}
\dim{\Xi(h,B)}=|T_{\varepsilon}(h)| \leq (n+1)^{d}2^{n(h+\varepsilon)}
\label{eqn:9}
\end{eqnarray}
\emph{Property2.2.} Given a mixed state $\rho$ with von Neumann entropy $S(\rho)=h$, if the eigenstates of $\rho $ lies in B, then for any fixed $\varepsilon > 0$ and $\delta > 0$, for sufficiently large n,
\begin{eqnarray}
tr(\Pi(h,B)\rho^{\otimes n})=\sum_{x^{n} \in T_{\varepsilon}(h)}p(x^{n}) \geq 1-\delta
\label{eqn:10}
\end{eqnarray}

Now let $\Upsilon(h)$ be the subspace of $\mathcal{H}^{\otimes n}$ which contains $\Xi(h,B)$ for all choices of basis $B$ and $\Pi_{\Upsilon}(h)$ be the projector onto $\Upsilon(h)$. Any other basis $B^{\prime}$ can be obtained from $B$ by applying some $d \times d$ unitary transformation $U$, thus $\Xi(h,B^{\prime})$ is obtained by applying $U^{\otimes n}$ to $\Xi(h,B)$. Then $\Upsilon(h)$ can be represented as the span of all $U^{\otimes n}\left |\phi \right \rangle$ where $U$ ranges over all $d \times d$ unitary matrices and $\left |\phi \right \rangle$ ranges over $\Xi(h,B)$.\\

\noindent\textbf{Definition3} \emph{The quantum entropy-typical subspace $\Upsilon(h)$ is defined as}
\begin{eqnarray}
\Upsilon(h)=span\{U^{\otimes n}|\phi\rangle:U \in \mathcal{U},|\phi\rangle \in \Xi(h,B) \}
\label{eqn:11}
\end{eqnarray}
\emph{where $\mathcal{U}$ is the collection of all $d \times d$ unitary matrixes.}\\

According to Ref\cite{Jozsa1998}, the expansion of dimension from $\Xi(h,B)$ to $\Upsilon(h)$ is up to $(n+1)^{d^{2}}$. Combined with property2.1, we have
\begin{eqnarray}
\dim{\Upsilon(h)} \leq (n+1)^{(d^{2}+d)}2^{n(h+\varepsilon)}
\label{eqn:12}
\end{eqnarray}

An immediate consequence of property2.2 is that $\Pi_{\Upsilon}(h)$ preserves $\rho^{\otimes n}$ approximately if $S(\rho)=h $. However, we give a stronger theorem below.\\

\noindent\textbf{Theorem1} \emph{Given a mixed state $\rho$, if the von Neumann entropy $S(\rho)\leq h $, then for any fixed $\varepsilon > 0$ and $\delta > 0$, for sufficiently large n,}
\begin{eqnarray}
tr(\Pi_{\Upsilon}(h)\rho^{\otimes n})\geq 1-\delta
\label{eqn:13}
\end{eqnarray}
\\
\textbf{Remark:} $\Pi_{\Upsilon}(h)$ preserves $\rho^{\otimes n}$ not only for the case that $S(\rho)=h$, but also for $S(\rho) < h$ !\\

To prove the theorem, we need the following lemma:\\

\noindent\textbf{Lemma1} \emph{Given a mixed state $\rho$, if $S(\rho)\leq h \leq d$, then there exist a basis $B^{\prime}=\{|e_{1}^{\prime}\rangle, |e _{2}^{\prime}\rangle... |e_{d}^{\prime} \rangle \}$ such that $S(\rho^{\prime})=h$, where $\rho^{\prime}=\sum_{i}\langle e_{i}^{\prime}|\rho|e_{i}^{\prime}\rangle|e_{i}^{\prime}\rangle\langle e_{i}^{\prime}|$.}\\

\noindent\emph{Proof}:
Suppose the spectrum decomposition of $\rho$ is $\rho=\sum_{k}p_{k}|e_{k}^{0}\rangle\langle e_{k}^{0}|$, where the eigenstates $|e_{k}^{0}\rangle $ lies in the basis $B^{0}=\{|e_{1}^{0}\rangle, |e _{2}^{0}\rangle... |e_{d}^{0} \rangle \}$. Define a basis $B^{1}=\left\{|e_{1}^{1}\rangle,|e_{2}^{1}\rangle...|e_{d}^{1}\rangle\right\}$ by
\begin{eqnarray}
|e_{l}^{1}\rangle=\frac{1}{\sqrt{d}}\sum_{k=1}^{d}\exp\left\{j2\pi\frac{kl}{d}\right\}  \left|e_{k}^{0}\right\rangle
\label{eqn:14}
\end{eqnarray}
where $j$ is the imaginary unit.
If we measure $\rho$ on the basis $B^{1}$, the result ensemble can be stated as $\rho^{1}=\sum_{l}\langle e^{1}_{l}|\rho|e^{1}_{l}\rangle|e^{1}_{l}\rangle\langle e^{1}_{l}|$. It can be verified that $S(\rho^{1})=d$.\\
Define an unitary operator $W$ by
\begin{eqnarray}
W=\sum_{i}|e_{i}^{1}\rangle\langle e_{i}^{0}|
\label{eqn:15}
\end{eqnarray}
Suppose the spectrum decomposition of $W$ is $W=\sum_{s}\exp\{j\theta_{s}\}|e^{W}_{s}\rangle\langle e^{W}_{s}|$.
With the basis $\{e_{s}^{W}\}$, we can define a function
\begin{eqnarray}
U(y_{1},y_{2},...,y_{d})=\sum_{s}\exp\{jy_{s}\}|e^{W}_{s}\rangle\langle e^{W}_{s}|
\label{eqn:16}
\end{eqnarray}
where $y_{s}\in [0,\theta_{s}]$. Obviously, $U$ is unitary, and
\begin{eqnarray}
U(0,0,...,0)=I, \ U(\theta_{1}, \theta_{2},...,\theta_{d})=W
\label{eq:17}
\end{eqnarray}
By applying $U(y_{1},y_{2},...,y_{d})$ on each state of the basis $B^{0}$, we can obtain the basis $B^{\mathbf{y}}=\{|e_{1}^{\mathbf{y}}\rangle, |e_{2}^{\mathbf{y}}\rangle,...|e_{d}^{\mathbf{y}}\rangle\}$, where $|e_{i}^{\mathbf{y}}\rangle=U(y_{1},y_{2},...,y_{d})|e_{i}^{0}\rangle$. If we measure $\rho$ on the basis $B^{\mathbf{y}}$, the result ensemble is
\begin{eqnarray}
\rho^{\mathbf{y}}&=&\sum_{i}\left\langle e_{i}^{\mathbf{y}}\right|\rho\left|e_{i}^{\mathbf{y}}\right\rangle \left|e_{i}^{\mathbf{y}}\right\rangle\left\langle e_{i}^{\mathbf{y}}\right| \nonumber\\
&=&\sum_{i}\left\langle e_{i}^{\mathbf{y}}\right|\left( \sum_{k}p_{k}\left|e_{k}^{0}\right\rangle\left\langle e_{k}^{0}\right|\right) \left|e_{i}^{\mathbf{y}}\right\rangle \left|e_{i}^{\mathbf{y}}\right\rangle\left\langle e_{i}^{\mathbf{y}}\right| \nonumber\\
&=&\sum_{i} \sum_{k}p_{k}\left|\langle e_{k}^{0}|e_{i}^{\mathbf{y}}\rangle\right|^{2}|e_{i}^{\mathbf{y}}\rangle\langle e_{i}^{\mathbf{y}}| \nonumber\\
&=&\sum_{i} \sum_{k}p_{k}\left|\langle e_{k}^{0}|U(y_{1},y_{2},...,y_{d})|e_{i}^{0}\rangle\right|^{2}|e_{i}^{\mathbf{y}}\rangle\langle e_{i}^{\mathbf{y}}| \nonumber\\
&=&\sum_{i} \sum_{k}p_{k}\left|\sum_{s}\exp\{jy_{s}\}\langle e_{k}^{0}|e_{s}^{W}\rangle\langle e_{s}^{W}|e_{i}^{0}\rangle\right|^{2} |e_{i}^{\mathbf{y}}\rangle\langle e_{i}^{\mathbf{y}}| \nonumber
\label{eqn:18}
\end{eqnarray}
Suppose the spectrum decomposition of $\rho^{\mathbf{y}}$ is $\rho^{\mathbf{y}}=\sum_{i}p_{i}^{\mathbf{y}}|e_{i}^{\mathbf{y}}\rangle\langle e_{i}^{\mathbf{y}}|$, then
\begin{eqnarray}
p_{i}^{\mathbf{y}}=\sum_{k}p_{k}\left|\sum_{s}\exp\{jy_{s}\}\langle e_{k}^{0}|e_{s}^{W}\rangle\langle e_{s}^{W}|e_{i}^{0}\rangle\right|^{2}
\label{eqn:19}
\end{eqnarray}
Since $S(\rho^{\mathbf{y}})=-\sum_{i}p_{i}^{\mathbf{y}}\log {p_{i}^{\mathbf{y}}}$, we can see that, $S(\rho^{\mathbf{y}})$ can be represented as a multi-variable function $S(y_{1},y_{2}...,y_{d})$ with domain $\left\{(y_{1},y_{2},...y_{d}) | \ 0\leq y_{s}\leq \theta_{s}, s=1,2,...d\right\}$. $S(y_{1},...,y_{d})$ is an elementary function, so it is continuous. Furthermore, it is easy to verify that
\begin{eqnarray}
S(0,0,...,0)=S(\rho), \ \ S(\theta_{1}, \theta_{2},...,\theta_{d})=d
\label{eq:20}
\end{eqnarray}
By the intermediate-value theorem, for any $h$ between $S(\rho)$ and $d$, there exist a point $(\alpha_{1},\alpha_{2},...\alpha_{d})$ in the domain such that $S(\alpha_{1},\alpha_{2},...,\alpha_{d})=h$.
Let $|e_{i}^{\prime}\rangle=U(\alpha_{1},\alpha_{2},...,\alpha_{d})|e_{i}^{0}\rangle$, then the basis $B^{\prime}=\{|e_{i}^{\prime}\rangle, i=1\ldots d\} $ is what we are finding.

With this lemma, we can prove theorem1.
\begin{eqnarray}
tr(\Pi_{\Upsilon}(h)\rho^{\otimes n})&\geq&tr[\Pi(h,B^{\prime})\rho^{\otimes n}]\nonumber \\
&=&\sum_{x^{n} \in T_{\varepsilon}(h)}\left\langle e_{x_{1}}^{\prime}e_{x_{2}}^{\prime}...e_{x_{n}}^{\prime}\right|\rho^{\otimes n}\left|e_{x_{1}}^{\prime}e_{x_{2}}^{\prime}...e_{x_{n}}^{\prime}\right\rangle \nonumber\\
&=&\sum_{x^{n} \in T_{\varepsilon}(h)}\prod_{i=1}^{n}\left\langle e_{x_{i}}^{\prime}\right|\rho\left|e_{x_{i}}^{\prime}\right\rangle \nonumber\\
&=&\sum_{x^{n} \in T_{\varepsilon}(h)}\prod_{i=1}^{n}\left\langle e_{x_{i}}^{\prime}\right|\rho^{\prime}\left|e_{x_{i}}^{\prime}\right\rangle \nonumber\\
&=&tr[\Pi(h,B^{\prime})\rho^{\prime \otimes n}]\geq 1-\delta \nonumber
\label{eqn:21}
\end{eqnarray}
The first inequality holds because $\Xi(h,B^{\prime})\subseteq \Upsilon(h)$.
The third equality holds because $\rho^{\prime}=\sum_{i}\left\langle e^{\prime}_{i}\right|\rho \left|e^{\prime}_{i}\right\rangle\left|e^{\prime}_{i}\right\rangle \left\langle e^{\prime}_{i}\right| $.
The last equality holds from (\ref{eqn:10}).

This result allows us to construct a universal compression scheme for all sources with von Neumman entropy $\leq h$ using the skill developed by Schumacher\cite{Schmacher1995,Wilde2011,Nielsen2000}.
More precisely, the encoding operation is the map $\mathcal{C}^{n}: \mathcal{H}^{\otimes n} \rightarrow \mathcal{H}^{n}_{c}$,
\begin{eqnarray}
\mathcal{C}^{n}(\sigma)\equiv \Pi_{\Upsilon}(h)\sigma\Pi_{\Upsilon}(h)+\sum_{u}E_{u}\sigma E_{u}^{\dag}
\label{eqn:22}
\end{eqnarray}
where $E_{u}\equiv |0\rangle\langle u|$. $|0\rangle$ is some standard state chosen from $\Upsilon(h)$ and $\left \{|u\rangle \right \} $ is an orthonormal basis for the orthocomplement of $\Upsilon(h) $.
The decoding operation is the map $\mathcal{D}^{n}: \mathcal{H}^{n}_{c} \rightarrow \mathcal{H}^{\otimes n}$, $\mathcal{D}^{n}(\sigma)\equiv\sigma$.
With the encoding and decoding operation, the fidelity of the compression
\begin{eqnarray}
F\left(\rho^{\otimes n},\mathcal{D}^{n}(\mathcal{C}^{n}(\rho^{\otimes n}))\right)&=&|tr(\Pi_{\Upsilon}(h)\rho^{\otimes n})|^{2}+\sum_{u}|tr(E_{u}\rho^{\otimes n})|^{2} \nonumber\\
&\geq & |tr(\Pi_{\Upsilon}(h)\rho^{\otimes n})|^{2} \nonumber\\
&\geq &|1-\delta|^{2}\geq 1-2\delta   \nonumber
\label{eqn:23}
\end{eqnarray}
$\delta$ can be made arbitrarily small for sufficiently large n, so the compression scheme is reliable.

The compression rate R is given by
\begin{eqnarray}
R&=&\lim_{n\rightarrow \infty} \frac{\log{\dim{\Upsilon(h)}}}{n} \nonumber \\
&\leq &\lim_{n\rightarrow\infty} (d^{2}+d)\frac{\log(n+1)}{n}+h+\varepsilon \nonumber
\label{eqn:24}
\end{eqnarray}
which tends  to $h+\varepsilon$. Because $\varepsilon$ can be as small as desired, the rate $h$ is achievable asymptotically. Thus we have shown that for any given $h$ and sufficiently large n, projection onto $\Upsilon(h)$ will provide a reliable compression for all sources with von Neumann entropy $\leq h$.

In this paper, we gives the definition of quantum entropy-typical subspace $\Upsilon(h)$. Then we show that any $\rho^{\otimes n}$ with $S(\rho)\leq h$ can be preserved approximately by $\Upsilon(h)$. This result implies a reliable universal compression scheme for the case that the von Neumann entropy of the source does not exceed $h$.\\

\begin{acknowledgments}
This work is supported by the National Natural Science Foundation of China Grant No.61271174.
\end{acknowledgments}
\providecommand{\noopsort}[1]{}\providecommand{\singleletter}[1]{#1}%
%

%\bibliography{paper2ref}

\end{document}